\documentstyle[aps, pre, amsfonts, amssymb, preprint, graphicx]{revtex}
\begin{document}
\draft
\author{S. Nazarenko, R. J. West and O. Zaboronski }
\address{Mathematics Institute, University of Warwick, Coventry CV4 7AL, 
United Kingdom}
\date{\today}
\title{Fourier space intermittency of the small-scale turbulent dynamo}
\maketitle

\begin{abstract}
The small-scale turbulent dynamo in the high Prandtl number regime
is described in terms of the one-point Fourier space correlators.
The second order correlator of this kind is the energy spectrum and
it has been previously studied in detail. We examine the higher order $k$-space
correlators which contain important information about the phases of the 
magnetic wavepackets and about the dominant  structures of the
magnetic turbulence which cause intermittency. In particular, the
fourth-order correlators contain information about the mean-square
phase difference between any two components of the magnetic field in a
plane transverse to the wavevector. This can be viewed as a measure 
of the magnetic field's polarization. Examining this new quantity, 
the magnetic field is shown to become plane polarized in the 
Kazantsev-Kraichnan model at large time, corresponding to a strong 
deviation from Gaussianity. We derive a closed equation for the 
generating function of the Fourier correlators and find the 
large-time asymptotic solutions of these correlators at all orders. 
The time scaling of these solutions implies the magnetic field has
log-normal statistics, whereas the wavenumber scaling indicates that the
field is dominated by intermittent fluctuations at high $k$.
\end{abstract}



\section{Introduction}
In some astrophysical applications such as the interstellar medium
and protogalactic plasmas the kinematic viscosity is greater than
the magnetic field diffusivity by the factor $10^{14}$ to $10^{22}$
\cite{kulsrud,schekoch}.
In these situations, there is a vast scaling range where the magnetic field has a
smaller characteristic length-scale than the velocity field.
The dynamo process of the stochastic stretching and amplification of 
the magnetic field can be studied in this case in terms of the statistics of Lagrangian
deformations \cite{chertkov,falkovich}. It can be pictured as a collection of 
magnetic field wavepackets, each moving along a fluid particle path and being 
distorted by the local strain.\footnote{The second and higher derivatives
of the velocity field can be ignored in this case.} Such a regime of smooth velocity 
fields is similar to the Batchelor's regime found in the related problem of passive scalar 
advection. The Batchelor's regime has a long history of study, some of the more 
recent advances can be found, for example, in \cite{schekoch,falkovich} 
(and references therein). A common further simplification of the problem is to assume
that the local strain matrix is a Gaussian white noise process, this is commonly known 
as the Kazantsev-Kraichnan model \cite{kazantsev,kraichnan}. However, this assumption is 
not always necessary, and some results have been shown to be universal for a broader 
class of stochastic flows \cite{chertkov}.

The first analysis of the dynamo problem were based in Fourier space
\cite{kazantsev,kraichnan,batchelor}, with particular emphasis being
placed on the second order moment, corresponding to the energy spectrum.
It was  realized, however, that the Fourier space and coordinate space 
equations have a similar structure. Recent studies of the turbulence 
intermittency (in both the passive scalar and dynamo problems) have 
focussed on coordinate space moments of second order or higher.
This approach was motivated by a feeling that this description is more 
natural and can give more information than the $k$-space moments 
\cite{chertkov,falkovich,schek2}. As a result, the only $k$-space correlator that 
has been seriously studied to date is the energy spectrum and there is
no theory describing the higher Fourier space correlators.

In the present paper, we turn our attention back to Fourier space
and consider the one-point moments of the magnetic field Fourier transform.
Although the problem under investigation here is the small-scale turbulent 
dynamo, we would like to give several reasons why the description of the 
Fourier space moments is important and why it should be developed in a 
broader turbulence context.

(i) The presence of singular structures in turbulence is known to
affect the scaling of the structure functions. The most well-known
examples of this being the $\beta$ and multi-fractal models \cite{Frisch}. 
However, some coherent structures are singular in Fourier space rather
than in coordinate space, and therefore can be detected by an investigation
of the Fourier moments. A simple example of structures that are singular 
in Fourier space (and regular in coordinate space) is a sea of vortex 
filaments in 2D turbulence. The layered pattern of the vortices in coordinate 
space corresponds to a 1D curve in Fourier space.\footnote{The Fourier transform should be 
taken over a local box, which is large enough to fit many layers of filaments, 
but is smaller than the large-scale vortices in this case. This will be discussed in the 
next section.}

(ii) To date, the only Fourier moment that was studied in detail is
the second order moment that describes the distribution of the turbulent
energy over wavenumbers. However, there exist other Fourier moments that have
a clear physical meaning and describe important properties of the turbulence.
One of these objects is related to the fourth order moments and has
a physical meaning of the mean turbulence polarization. It will be
introduced and studied in this paper.

(iii) Phases of the Fourier modes can be dealt with directly, and
therefore the validity of the random phase assumption can be examined.

(iv) Finally, in some cases a Fourier space analysis is the only way to
have a treatable problem due to the introduction of non-localities, for example
via a pressure or wave dispersion. Note that the dynamo and passive
scalar systems do not fit into this class of problems. However, the methods 
developed in the present paper will be applied to the non-local Navier-Stokes
equation\footnote{Such an equation is the 
basis of Rapid Distortion theory which will be developed for the case of a 
stochastic strain.} (which involves a pressure term) in our next paper \cite{NazZab}.

\section{One-point and two-point correlators of Fourier amplitudes}
To avoid any confusion, we should state that the Fourier transforms in
this paper are performed over a finite box of a size much greater than typical
length-scale of the magnetic turbulence, but much less than the scale of the advecting
velocity field. It is only for finite box Fourier transforms that one-point correlators are 
well defined objects. Secondly, the center of each finite box is not fixed. The box center coordinate
dependence of the Fourier transforms is a useful measure of the slow variability of the large-scale
magnetic turbulence. In this paper we make each box move in unison with a fluid particle, and thus
any coordinate dependence is replaced by a time dependence along a given fluid trajectory. 

One should be aware that in this case the finite box size does not produce a discrete spectrum as the 
magnetic field is in general not periodic within the box. On the other hand, the 
lack of periodicity can produce unphysical power-law tails in the Fourier transforms at (large)
wavenumbers that are greater than one over the box length. Slow decay of these tails can be undesirable in
numerical simulations and other applications. However, this problem can be easily cured by
using a smooth filter instead of the box, for example by using Gabor transforms instead \cite{NKD}. 
In this presentation the box (or filter) shape is actually not an issue. We will present the 
results in terms of finite box Fourier transforms. However, performing the analysis using
Gabor transforms will in fact produce identical results.

Let us consider the general two-point correlator of the Fourier transformed magnetic field 
components
\begin{equation}
 \langle B_{i_1}({\mathbf k}_1)  B_{i_2}({\mathbf k}_1) \cdots  B_{i_n}({\mathbf k}_1) 
 B_{j_1}({\mathbf k}_2)  B_{j_2}({\mathbf k}_2) \cdots  B_{j_n}({\mathbf k}_2)  \rangle,
\end{equation}
where $i_1, i_2, \cdots , i_n$ and $j_1, j_2, \cdots , j_n$ take the values 1, 2 or 3 corresponding 
to the three components of the magnetic field, and $n$ is an arbitrary natural number.
Let us assume the turbulence is quasi-homogeneous. That is, the correlator
$\langle B_{i}({\mathbf x}_1) B_{j}({\mathbf x}_2) \rangle $ (and similar)
depend only on ${\mathbf x}_1 + {\mathbf x}_2 $ as slowly as the advecting strain field.
On the other hand, such correlators decay in ${\mathbf x}_1 - {\mathbf x}_2 $
at distances much less than the box size $L$. In this case
\begin{eqnarray}
 \langle B_{i_1}({\mathbf k}_1) && B_{i_2}({\mathbf k}_1) \cdots  B_{i_n}({\mathbf k}_1) 
 B_{j_1}({\mathbf k}_2)  B_{j_2}({\mathbf k}_2) \cdots  B_{j_n}({\mathbf k}_2)  \rangle
 =  \\
 && \langle B_{i_1}({\mathbf k}_1)  B_{i_2}({\mathbf k}_1) \cdots  B_{i_n}({\mathbf k}_1) 
 B_{j_1}(-{\mathbf k}_1)  B_{j_2}(-{\mathbf k}_1) \cdots  B_{j_n}(-{\mathbf k}_1) \rangle \;
 f({\mathbf k}_1 + {\mathbf k}_2),\nonumber
\end{eqnarray}
where $f({\mathbf k}) = \frac{1}{L^d} \Pi_{l=1}^d \frac{\sin(L k_{l})}{k_{l}} $ 
is the Fourier transform of the filter function and $d$ is the number of space dimensions. 
Here we see that any two-point correlator is fully determined
by the one-point correlators
\begin{equation}
 I_{i_1, i_2, \cdots , i_n;j_1, j_2, \cdots , j_n}({\mathbf k}) =
 \langle B_{i_1}({\mathbf k})  B_{i_2}({\mathbf k}) \cdots  B_{i_n}({\mathbf k}) 
 B_{j_1}(-{\mathbf k})  B_{j_2}(-{\mathbf k}) \cdots  B_{j_n}(-{\mathbf k}) \rangle.
\end{equation}
This simple observation means we can avoid lengthy derivations dealing
with two-point objects directly, and instead can obtain results from the 
one-point correlators that are themselves far easier to deal with.
Turbulence isotropy and the divergence-free condition further narrow 
the class of possible one-point (and two-point) correlator tensors
we need to consider (see for example \cite{McComb})
\begin{equation}
 I_{i_1, i_2, \cdots , i_n;j_1, j_2, \cdots , j_n}({\mathbf k}) =
 \sum_{\rm pairs}
 \Phi^n_s (k) 
 D_{l_1, l_2}
 D_{l_3, l_4} \cdots
 D_{l_{n-1}, l_n},
 \label{1point}
\end{equation}
where
\begin{equation}
   D_{ij} \equiv \delta_{i,j} - \frac{k_i k_j}{k^2},
\end{equation}
and the summation is over the set of all possible pairings
$(l_1, l_2), (l_3,l_4), \cdots ,(l_{n-1},l_n)$ of the indices $i$ and $j$
from the left hand-side of (\ref{1point}). For each pairing, the index 
$s$ is equal to the number of pairs that consist only of $i$'s,
\footnote{ There are obviously going to be the same number of pairs 
that consist only of $j$'s.} for example $(i_1, i_3)$ or  $(i_8,i_2)$. 
One can see that any correlator of order $2n$ can be expressed in terms of
$N=\hbox{Max} (s) +1=\hbox{Int}(n/2)+1$ independent functions
$\Phi^n_s (k)$. There is only one such function for
the second order correlators, two functions for the fourth 
and sixth orders, and three functions at the eighth 
and tenth orders.

Instead of $\Phi^n_s (k)$, we can choose another set of independent
functions, in particular the following set of correlators
\begin{equation}
 \Psi^n_s =
 \langle |{\mathbf B}({\mathbf k})|^{(2n-4s)}  |{\mathbf B}({\mathbf k})^2|^{2s} \rangle.
 \label{basis}
\end{equation}
These correlators are  linear combinations of $\Phi^n_s (k)$'s,
for example at the fourth order we have
\begin{eqnarray}
 \Psi^{(2)}_0 &=& d(d-1) \Phi^{(2)}_0 + (d-1) \Phi^{(2)}_1, \nonumber \\
 \Psi^{(2)}_1 &=& 2 (d-1) \Phi^{(2)}_0 + (d-1)^2 \Phi^{(2)}_1. \nonumber 
\end{eqnarray}
Below, we will derive a technique that allows us to deal with
the fundamental set of correlators (\ref{basis}).

%
\section{The model}
%
%
Let us start with the equation for the magnetic field ${\mathbf B}({\mathbf x}, t)$,
\begin{equation}
 \partial_t {\mathbf B} + ({\mathbf u} \cdot \nabla)  {\mathbf B} =
 ({\mathbf B} \cdot \nabla)  {\mathbf u} + \kappa \Delta  {\mathbf B},
 \label{mhd}
\end{equation}
where ${\mathbf u}$ is the velocity field, which we assume has a much slower spatial
variation than the magnetic field ${\mathbf B}$, and $\kappa $ is the magnetic diffusivity
(determined by the fluid conductivity). We will make use of the previously discussed
finite box Fourier transforms. Each box has sides of length $L$ which is chosen to lie 
between the length-scales associated with ${\mathbf B}$ and ${\mathbf u}$, namely $L_{B}$ and $L_{u}$.
The box Fourier transform is defined as
\begin{equation}
 \hat{{\mathbf B}}({\mathbf k},{\mathbf x},t) \int_{box} {\mathbf B}({\mathbf r},t) 
 \, e^{ i{\mathbf k} \cdot ({\mathbf x}-{\mathbf r})} \, d {\mathbf r},
\end{equation}
where ${\mathbf x}$ is the coordinate of the box center.
Applying this Fourier transform to (\ref{mhd}) we have
(with accuracy up to the first order of the scale separation
parameter $\epsilon = L_{B}/L_{u} \ll 1$) 
\begin{equation}
  \partial_t \hat B_m  + u_i  \nabla_i \hat B_m = \sigma_{ij} k_i \, \partial_j
  \hat B_m + \sigma_{mi} \hat B_i - \kappa k^2 \hat B_m,
\end{equation}
where $\sigma_{ij} = \nabla_j u_i$ is the strain matrix and the operators
$\nabla_i$ and $\partial_i$ correspond to derivatives with respect to $x_i$ and $k_i$
respectively. In considering a fluid path determined by $\dot{{\mathbf x}}(t) ={\mathbf u}$, 
and thus noting that $\sigma_{ij} ( {\mathbf x}, t) \to \sigma_{ij} ( {\mathbf x}(t), t)$ and 
$\hat{{\mathbf B}}({\mathbf k},{\mathbf x},t) \to \hat{{\mathbf B}}({\mathbf k},{\mathbf x}(t),t)$,
this equation becomes\footnote{Hereafter, we will the drop hat notation on $\hat{{\mathbf B}}$
because only Fourier components will be considered. Also, we will not mention
explicitly dependence on the fluid path and simply write ${\mathbf B} \equiv {\mathbf B} ({\mathbf k}, t)$.}
\begin{equation}
 \partial_t B_m  = \sigma_{ij} k_i \, \partial_j
 B_m + \sigma_{mi} B_i - \kappa k^2 B_m.
 \label{beqn}
\end{equation}
It should also be noted that the strain $\sigma_{ij}$, taken along a fluid path, 
only enters this equation as a given function of time. To complete the model one has to 
specify this dependence or postulate the strain statistics. In what follows we choose the 
strain matrix to be Gaussian such that 
\begin{equation}
 \sigma_{ij} = \Omega \left(A_{ij} - \frac{A_{ll}}{d} \delta_{ij}\right),
\end{equation}
where $A_{ij}$ is a matrix with statistically independent elements that are white in time
\begin{equation}
 \langle A_{ij}(t) A_{kl}(0) \rangle = \delta_{ij} \delta_{kl}  \delta(t).
\end{equation}
This choice of strain matrix ensures incompressibility and statistical isotropy.
Indeed, 
\begin{equation}
 \langle \sigma_{ij}(t) \sigma_{kl}(0) \rangle = \Omega \left(\delta_{ik} \delta_{jl} -
\frac{1}{d} \delta_{ij} \delta_{kl}\right)  \delta(t).
\end{equation}
However, one must be aware that this is not the only way to satisfy the incompressibility 
and isotropy conditions; for example \cite{chertkov,falkovich,balkovski} have chosen
\begin{equation}
  \langle \sigma_{ij}(t) \sigma_{kl}(0) \rangle = \Omega \left[(d+1) \delta_{ik} \delta_{jl} -
  \delta_{il} \delta_{jk} -\delta_{ij} \delta_{kl}\right]  \delta(t).
\end{equation}
More generally, there is an infinite one-parametric family of possible strains.
The strain matrices that are Gaussian and white in time correspond to the 
Kazantsev-Kraichnan model, a natural starting point for an analytical analysis
because of its simplicity. However, it should be noted that some results remain 
universal in the case of other smooth velocity fields found in a 
much wider class of statistical models \cite{chertkov,falkovich,balkovski}.
In our future work we will investigate the behaviour of the 
Fourier space correlators in the case of more general statistics.

%
\section{Generating Function}
%
%
Let us consider the following generating function
\begin{equation}
  Z(\lambda, \alpha, \beta, k) = \langle e^{\lambda|{\mathbf B}({\mathbf k})|^2 
 + \alpha {\mathbf B}^2({\mathbf k}) + \beta \overline{{\mathbf B}}^2({\mathbf k})}
  \rangle ,
 \label{z} 
\end{equation}
where the overline denotes the complex conjugation. This function allows one
to obtain any of the fundamental one-point correlators (\ref{basis}) via differentiation
with respect to $\lambda, \alpha$ and $ \beta$
\begin{equation}
  \Psi^n_s =
  \langle |{\mathbf B}({\mathbf k})|^{(2n-4s)}  |{\mathbf B}^2({\mathbf k})|^{2s} \rangle   
  = \left[ \partial_\lambda^{(2n-4s)}   \partial_\alpha ^{s}    \partial_\beta ^{s}  Z  
  \right]_{\lambda= \alpha= \beta=0} .
  \label{psi_v_z}
\end{equation}
Differentiating (\ref{z}) with respect to time and using
the dynamical equation  (\ref{beqn}), we have
\begin{eqnarray}\label{zdot}
 \dot Z =  k_i \partial_j
 \langle \sigma_{ij} E \rangle +
 && \lambda \langle \sigma_{ml} ({\overline B_m} B_l + {\overline B_l} B_m) E
 \rangle + 2 \alpha  \langle \sigma_{ml} { B_m} B_l E \rangle
 + 2 \beta  \langle \sigma_{ml} {\overline B_m} {\overline B_l} E \rangle 
 \\
 && - 2 \kappa k^2 \langle [{\lambda|{\mathbf B}({\mathbf k})}|^2 
 + \alpha {\mathbf B}^2({\mathbf k}) + \beta \overline{{\mathbf B}}^2({\mathbf k})] E
 \rangle,\nonumber
\end{eqnarray}
where 
\begin{equation}
 E = e^{\lambda|{\mathbf B}({\mathbf k})|^2 
 + \alpha {\mathbf B}^2({\mathbf k}) + \beta \overline{{\mathbf B}}^2({\mathbf k})} .
 \label{E} 
\end{equation}
To find the correlators on the right hand-side of (\ref{zdot}), one needs to make
use of the Gaussianity of the strain matrix $\sigma_{ij}$ and 
perform a Gaussian integration by parts. 
We then use the whiteness of the strain field to find the response function
(functional derivative of $B_l$ with respect to $\sigma_{ij}$).
Finally, one can use the statistical isotropy of the strain, so that the final equation
involves only $k = |{\mathbf k}|$ and no angular dependence of the wavevector.
This derivation is discussed in more detail in the Appendix. Here, we just write the 
final result
\begin{eqnarray}  \label{zdotF}
 \dot Z &=& \frac{\Omega}{2} \left[ \left(1 - \frac{1}{d}\right) k^2 Z_{kk} + 
 \frac{1}{d} (-4 {\cal D} + d^2 -1) k Z_{k} \right. \\
 && \left. +(2d-6) {\cal D} Z + 4\left(1-\frac{1}{d}\right) {\cal D}^2 Z
  + 2 (\lambda^2 -4 \alpha \beta) (Z_{\alpha \beta}  - Z_{\lambda \lambda})
 \right]
 -2 \kappa k^2 {\cal D} Z,
 \nonumber
\end{eqnarray}
where the $k, \alpha, \beta$ and $\lambda$ subscripts in $Z$ denote 
differentiation with respect to $k, \alpha, \beta$ and $\lambda$
respectively, and
\begin{equation}
 {\cal D} = \lambda \partial_\lambda + \alpha \partial_\alpha +
 \beta \partial_\beta.
 \label{calD} 
\end{equation}
Actually, one can reduce the number of independent variables in this equation
by taking into account that $Z$ depends on
$\alpha$ and $\beta$ only in the combination $\eta = \alpha \beta$.
The easiest way to see this is to consider a Taylor series expansion
of $Z$ from (\ref{z}); any term that contains
 $\alpha$ and $\beta$ in a different combination will be zero because
of the quasi-homogeneity of the turbulence. Thus, we can write
\begin{eqnarray}\label{zdotFF1}
 \dot Z &=& \frac{\Omega}{2} \left[ \left(1 - \frac{1}{d}\right) k^2 Z_{kk} + 
 \frac{1}{d} (-4 {\cal D} + d^2 -1) k Z_{k} \right. \\
 && \left. +(2d-6) {\cal D} Z + 4\left(1-\frac{1}{d}\right) {\cal D}^2 Z
  + 2 (\lambda^2 -4 \eta) (Z_{\eta} + 
 \eta Z_{\eta \eta}  - Z_{\lambda \lambda})
 \right]
 -2 \kappa k^2 {\cal D} Z,
 \nonumber
\end{eqnarray}
where 
\begin{equation}
 {\cal D} = \lambda \partial_\lambda + 2 \eta \partial_\eta.
 \label{calD1} 
\end{equation}
In what follows, we will restrict our consideration to the 3D case
($d=3$) in which case the equation for $Z$ reduces to
\begin{eqnarray}\label{zdotFF}
 \dot Z &=& \frac{\Omega}{3} \left[ k^2 Z_{kk} + 
 (4 - 2 {\cal D}) k Z_{k} \right. \\
 && \left. + 4 {\cal D}^2 Z
  + 3 (\lambda^2 -4 \eta) (Z_{\eta} + 
 \eta Z_{\eta \eta}  - Z_{\lambda \lambda})
 \right]
 -2 \kappa k^2 {\cal D} Z.
 \nonumber
\end{eqnarray}
%

%
\section{Energy spectrum}
%
%
Let us now consider the energy spectrum of the magnetic turbulence,
given by the second order correlator
\begin{equation}
 E(k,t) = \Psi^1_0 =
 \langle |{\mathbf B}({\mathbf k})|^{2}   \rangle   = 
 \left[ \partial_\lambda  Z  
 \right]_{\lambda= \eta=0} .
 \label{spectrum}
\end{equation}
Differentiating (\ref{zdotFF}) with respect to $\lambda$ and taking the result
at $\lambda= \eta=0$ we have
\begin{equation}
 \dot E = \frac{\Omega}{3} (k^2 E_{kk} + 2 k E_{k} + 4 E) - 2 \kappa k^2 E.
 \label{Eeqn}
\end{equation}
This equation for the evolution of the energy spectrum was first obtained by
Kazantsev \cite{kazantsev} and Kraichnan and Nagarajan \cite{kraichnan}.
Kazantsev analyzed an eigenvalue problem associated with this equation
which allowed him to obtain the growth exponents of the total magnetic energy.
Numerically, the energy spectrum was studied by Kulsrud and Anderson \cite{kulsrud}
who gave a detailed description of the $k$-space evolution of this spectrum.
Recently, Schekochihin, Boldyrev and Kulsrud \cite{schekoch}
presented the complete solution of (\ref{Eeqn}) obtained by the use of 
Kontorovich-Lebedev transforms (KLT). Note that this integral transform approach has 
an advantage over the Kazantsev's eigenvalue analysis in that it allows us to obtain 
not only the growth exponents, but also the power-law prefactors, of the large time 
asymptotic solutions. 

As they will be of use later in this presentation, let us briefly review the previous 
results for the energy spectrum before we discuss the higher order moments.
Using the following substitution \cite{schekoch}
\begin{equation}
 E  = e^{5  \Omega t/4} \, 
 k^{-1/2} \, \phi(k/k_{d_1} , t) \;\;\; k_{d_n} = \sqrt{\frac{\Omega}{6n \kappa}},
 \label{univ}
\end{equation}
one can reduce (\ref{Eeqn}) to
\begin{equation}
 \frac{3}{\Omega} \dot \phi(p,t) =  p^2 \phi_{pp} + 
  p \phi_p -p^2 \phi,
 \label{phidot}
\end{equation}
where $p=k/k_{d_1}$. At scales much greater than the dissipative one, $p \ll 1$, there
is a perfect conductor regime for time $t \ll (\ln q)^2$ (where $q \ll 1$ is the 
mean wavenumber of the initial condition). Thus, in this regime the last term 
in (\ref{Eeqn}) can be neglected. By changing to logarithmic coordinates and a 
moving frame of reference one can transform this equation into a heat equation. 
For $t \gg 1/\Omega$ the solution of which is just the Green's function, 
which gives \cite{kulsrud,schekoch,robsPhD,robsPaper}
\begin{equation}
 \phi  = \hbox{const} \, t^{-1/2} \, 
 e^{ - \frac{3 (\ln k/q)^2}{4 \Omega t}},
 \label{phisoln}
\end{equation}
where the constant is fixed by the initial condition. This solution 
describes a spectrum with an expanding $k^{-1/2}$
scaling range. At $t \sim (\ln q)^2$ the front of this scaling range reaches 
the dissipative scales. To solve (\ref{Eeqn}) in this case, we note
that the right hand-side of this equation is just the modified Bessel operator
and by using KLT one immediately obtains 
\cite{schekoch}
\begin{equation}
 \phi(p,t) =  \hbox{const} 
 \int_0^\infty ds \, s \, \hbox{sinh} (\pi s) \, K_{is} (p)  K_{is} (q) \,
 e^{-s^2 t},
 \label{phidis}
\end{equation}
where $K_{is}$ is a MacDonald function of imaginary order.
Again this solution is given by the Green's function only because
the condition that $t \gg 1/\Omega$ is obviously satisfied if $q \ll 1$.
For time $t \gg (\ln q)^2$, the function $e^{-s^2 t}$ is strongly peaked
at $t=0$ and integration of (\ref{phidis}) gives
\begin{equation}
 \phi =  \hbox{const} \,\,\,t^{-3/2} K_0(p).
 \label{phiinf}
\end{equation}
One should note that although the $K_0(p)$ shape is somehow predicted by the
Kazantsev's eigenmode analysis \cite{schekoch,kazantsev,robsPhD,robsPaper}, 
the $t^{3/2}$ factor can only be obtained by solving the full initial value
problem. For $p \ll 1$, $K_0(p) \approx - \ln p$, which means that
at large times the scales far larger than the dissipative one are
affected by diffusion via a logarithmic correction
\begin{equation}
 E(k) \sim k^{-1/2} \, \ln (k_{d_1}/k).
 \label{logcor}
\end{equation}
The energy evolution gives an important, but incomplete picture, of the 
dynamo process. In particular, it does not capture the existence of 
small-scale intermittency and does not allow us to predict the type of 
coherent structures dominating the turbulence at large time. To deal with 
these issues one has to study higher order correlators. Higher one-point 
correlators in coordinate space were studied in \cite{chertkov} and used 
to predict the shape of the dominant structures. Below, we proceed 
to study the higher $k$-space correlators. In particular, this will lead 
to the discovery of a new quantity of interest, corresponding to the
mean polarization of the magnetic turbulence.

\section{4th-order correlators, turbulence polarization and flatness}
%
There are two independent 4th order correlators 
\begin{mathletters}
\label{SandT}
\begin{equation}
 S(k,t) = \Psi^{(2)}_0 =
 \langle |{\mathbf B}({\mathbf k})|^{4}   \rangle   
 = \left[   Z_{\lambda \lambda} 
 \right]_{\lambda = \eta=0},
\end{equation}
\begin{equation}
 T(k,t) = \Psi^{(2)}_1 =
 \langle |{\mathbf B}^2({\mathbf k})|^{2}   \rangle    
 = \left[   Z_{\eta} 
 \right]_{\lambda= \eta=0} .
\end{equation}
\end{mathletters}
Differentiating (\ref{zdotFF}) twice with respect to $\lambda$ and taking the result
at $\lambda= \eta=0$ we have
\begin{equation}
 \dot S = \frac{\Omega}{3} (k^2 S_{kk} + 10 S + 6 T) - 4 \kappa k^2 S.
 \label{Seqn}
\end{equation}
Now, differentiating (\ref{zdotFF}) with respect to $\eta$ and taking the result
at $\lambda= \eta=0$ we get
\begin{equation}
 \dot T = \frac{\Omega}{3} (k^2 T_{kk} + 4 T +12 S) - 4 \kappa k^2 T.
 \label{Teqn}
\end{equation}
Equations (\ref{Seqn}) and (\ref{Teqn}) make up a complete system for $S$ and $T$ and can be 
solved exactly in the general case. Observe that there is a closed equation
for $W=S-T$
\begin{equation}
 \dot W = \frac{\Omega}{3} (k^2 W_{kk} - 2 W) - 4 \kappa k^2 W.
 \label{Weqn}
\end{equation}
Before solving this equation, let us examine the physical meaning of $W$ by writing
it as 
\begin{equation}
 W = 4 \sum_{j \ne l}^3 \langle [\Im (B_j {\overline B_l} )]^2 \rangle =
 4 \sum_{j \ne l}^3 \langle |B_j|^2 |B_l|^2  \sin^2 (\phi_j-\phi_l) \rangle \ge 0,
 \label{W}
\end{equation}
where $\Im$ denotes the imaginary part and $\phi_j$ and $\phi_l$ are the phases 
of the components $B_j$ and $B_l$ respectively. We see therefore that $W$ contains 
information not only about the amplitudes but also about the phases of the Fourier modes.
In particular,  $W \equiv 0$ corresponds to the case where all Fourier components of
the magnetic field are plane polarization. If $W \ne 0$ then other polarizations
(circular, elliptic) are present. This is the case for example for a Gaussian field
where one finds $W= E^2 /2 >0$. On the other hand, the smallness of the phase differences in $W$ 
can be overpowered by large amplitudes. Therefore, a better measure of the mean
polarization would be a normalized $W$, for example
\begin{equation}
 P= W/S.
 \label{P}
\end{equation}
Defined in this way, the mean turbulence polarization is an example of an important
physical quantity which can be obtained from the one-point Fourier correlators, 
and that is unavailable from the (one-point or two-point) coordinate space correlators.

In the perfect conductor regime, when the diffusivity term in
the equations (\ref{Seqn}) -- (\ref{Weqn}) can be ignored, the solution
for $W$ can be obtained in a similar manner to the previous energy spectrum analysis. 
Namely, by reducing (\ref{Weqn}) into a heat equation via a logarithmic change of 
variables and passing into a moving frame of reference. The solution therefore is
\begin{equation}
 W = W_0 \, t^{-1/2} e^{-3 \Omega t /4} \, k^{1/2} e^{ - \frac{3 (\ln k/q)^2}{4 \Omega t}},
 \label{Wsoln}
\end{equation}
where $W_0$ is a constant which can be found from the initial conditions.
We see that  $W$ develops a $ k^{1/2}$ scaling range which is cut-off at low and
high $k$ by exponentially propagating fronts. Within this scaling range, 
$W$ decays exponentially in time.

Given $W$, one can also find $S$ by representing it as $S=V+cW$ and choosing the constant $c$
such that the equation for $V$ is closed. This gives $c=3/7$ and
\begin{eqnarray}
 S &=&  t^{-1/2}  \, k^{1/2} e^{ - \frac{3 (\ln k/q)^2}{4 \Omega t}}
 \left( V_0 \, e^{21\Omega t /4} + \frac{3}{7} W_0 \,  e^{-3 \Omega t /4} \right),
 \label{Ssoln}
\end{eqnarray}
where $V_0$ is another constant which can be found from the initial conditions.
For $t \gg 1/\Omega$, the second term in the parenthesis should be neglected, and
we have the following solution for the mean turbulence polarization
\begin{equation}
 P =  W/S = \frac{W_0}{V_0} e^{- 6\Omega t}.
 \label{Psoln}
\end{equation}
In the perfect conductor regime we see the mean polarization
tends to a value that is independent of $k$ and decays exponentially in time.
This means that all Fourier modes of the magnetic field eventually become 
plane polarized. Recall that such turbulence is very far from being Gaussian,
remembering that the mean polarization of a Gaussian field remains finite.\footnote{That 
is, elliptic and circular polarized modes are present.} 

It is also easy to obtain solutions to $W$ for the diffusive regime.
Indeed, following the example of the energy spectrum we make use of the substitutions
\begin{eqnarray}
   W &=& e^{-3\Omega t /4} k^{1/2} \phi (k/k_{d_2}),\\
   V &=& e^{21\Omega t /4} k^{1/2} \phi (k/k_{d_2}),
\end{eqnarray}
to transform the governing equations for $W$ and $V$ into 
a form similar to (\ref{phidot}). In each case we can solve the equation 
for $\phi$ using KLT and find that for times $t \gg (\ln q)^2$ 
\begin{eqnarray}
   W &\simeq& W_0 k^{1/2} t^{-3/2} e^{-3\Omega t /4} K_0(k/k_{d_2}),\\
   V &\simeq& V_0 k^{1/2} t^{-3/2} e^{21\Omega t /4} K_0(k/k_{d_2}).
\end{eqnarray}
Correspondingly, $S$ has the solution
\begin{equation}
S \simeq k^{1/2}t^{-3/2}\left[V_0e^{21\Omega t/4}
                       +\frac{3}{7}W_0e^{-3\Omega t/4}\right]
                         K_0 (k/k_{d_2} ),
\end{equation}
while importantly we find that the normalised polarisation $P=W/S$ behaves identically in both 
the diffusive and perfect conductor regimes.

Therefore, in the diffusive regime $W$ continues to decrease 
in time with the same exponential rate as perfect conductor case
and has the same $t^{-3/2}$ prefactor as the energy spectrum. 
Thus, by the time the diffusive regime is achieved $W$ can be essentially put equal to zero.
The fact that the polarization becomes plane has quite a simple physical explanation.
Indeed, a magnetic field wavepacket of arbitrary polarization will be strongly
distorted by stretching. The stretching being strongest along the direction of 
the dominant eigenvector of the Lagrangian deformation matrix 
(corresponding to the greatest Lyapunov exponent). Such a stretching will make any 
initial ``spiral'' structure flat at large time, with the
dominant field component lying in a plane passing through the eigenvector stretching 
and wavevector directions.\footnote{Of course, the incompressibility of ${\mathbf B} ({\mathbf k})$ ensures that it 
is perpendicular to ${\mathbf k}$.}

Another measure of intermittency in turbulence is the flatness which
can be defined in $k$-space as $F=S/E^2$. For large time, in the
perfect conductor regime
\begin{equation}
 F \sim  t^{1/2} \, e^{11 \Omega t/4} \, k^{3/2}.
 \label{Fsoln}
\end{equation}
We see that the flatness grows both in time and in $k$ which
indicates the presence of small-scale intermittency. Such an intermittency
can be attributed to the presence of coherent structures in $k$-space.
In the diffusive regime, in contrast to the polarisation
the behaviour of the flatness is modified and at large time one finds
\begin{equation}\label{Fsol2}
   F \simeq k^{3/2} t^{3/2} e^{11\Omega t/4} \frac{K_0(k/k_{d_2})}{[K_0(k/k_{d_1})]^2}.
\end{equation}
For small $k$, we again have a region of $k^{3/2}$ scaling but now with  
a log correction arising from the MacDonald functions. For large $k$, below the spectral cut-off,
the additional MacDonald functions act to heighten the flatness. That is, the introduction of 
a finite diffusivity actually increases the small-scale intermittency.
In the next section we will investigate the correlators of all orders.

\section{Large-time behavior of higher correlators}
%
The observation in the last section, that there is a dominant
field component, allows us to predict that for large time
$|{\mathbf B}|^4 \approx |{\mathbf B}^2|^2 $ in each realization, that is 
$ Z_{\lambda \lambda} \approx Z_{\alpha \beta}$. 
Therefore,  the property $ Z_{\lambda \lambda} = Z_{\alpha \beta}$, if valid initially, 
should be preserved by the equation for $Z$. Indeed, let us differentiate
the equation (\ref{zdotFF}) twice with respect to $\lambda$ and
subtract it from the same equation differentiated with respect to 
$\alpha$ and $\beta$. This gives the following closed equation
for the combination $w= Z_{\lambda \lambda} - Z_{\alpha \beta}$
\begin{eqnarray} \label{wdot}
 \dot w &=& \frac{\Omega}{3} \left[ k^2 w_{kk} - 
 2 k {\cal D} w_{k}  + 4 {\cal D}^2 w \right. \\
 && \left.  + 16 ( {\cal D}w + w)
  + 3 (\partial_{\lambda \lambda} - \partial_{\alpha \beta})
 \{ (\lambda^2 -4 \alpha \beta ) w \}
 \right]
 -2 \kappa k^2 ({\cal D} w +2 w).
 \nonumber
\end{eqnarray}
We see that if $w \equiv 0$ at $t=0$, then it will remain identically
zero for all time. Thus, we can consider a class of solutions of
(\ref{zdotFF}), corresponding to large-time asymptotics of the general
solution, such that $ Z_{\lambda \lambda} = Z_{\alpha \beta}$.
Assuming this equality in (\ref{zdotFF}) and putting $\eta=0$ we have
\begin{equation}
 \dot Z = \frac{\Omega}{3} \big[  k^2 Z_{kk} + 
 (-2 \lambda \partial_\lambda + 4) k Z_{k}  
 + 4 (\lambda \partial_\lambda)^2 Z
 \big]
 -2 \kappa k^2 \lambda \partial_\lambda Z.
 \label{zdotDeg}
\end{equation}
Let us consider a solution to this equation that is formally 
represented as a series in $\lambda$ (for example via a Taylor series):
\begin{equation}
 Z = 1 +  \sum_{n=1}^{\infty} \frac{\lambda^n}{n!} \, \Psi^{(n)}(k, t).
 \label{taylorZ}
\end{equation}
Here, the function $\Psi^{(n)}$ is the correlator of order $2n$.
We omitted the lower index in $\Psi^{(n)}$ because the correlators
corresponding to different lower indices are identical
in this case. Substituting (\ref{taylorZ}) into  (\ref{zdotDeg})
we have the following equation for these correlators
\begin{equation}
 \dot \Psi^{(n)} = \frac{\Omega}{3} \big[  k^2 \Psi^{(n)}_{kk} + 
 (-2 n + 4) k \Psi^{(n)}_{k}  
 + 4 n^2 \Psi^{(n)}
 \big]
 -2 \kappa k^2 n \Psi^{(n)}.
 \label{zdotDeg1}
\end{equation}
Note that for $n=1$ this equation agrees with the Kazantsev equation 
for the energy spectrum (\ref{Eeqn}). Moreover, by the substitution
\begin{equation}
 \Psi^{(n)}  = e^{(n-1/2)(n+3/2)\Omega t} \,
 k^{n-3/2} \, \phi(k/k_{d_n} , t), 
 \label{univ1}
\end{equation}
one can reduce (\ref{zdotDeg1}) to an equation for $\phi$, similar to (\ref{phidot}), 
that is independent of $n$. We can therefore immediately write down the general 
solution for the correlators at any order $n$. In particular, in the perfect conductor
regime $1/\Omega \ll t \ll (\ln q)^2$ we have
\begin{equation}
 \Psi^{(n)}  = \hbox{const}(n) 
 \, t^{-1/2} \, e^{(n-1/2)(n+3/2)\Omega t}
 \, e^{- \frac{3 (\ln k/q)^2}{4 \Omega t}} \, 
 k^{n-3/2},
 \label{Psi_n_pc}
\end{equation}
and in the diffusive regime $t \gg (\ln q)^2$
\begin{equation}
 \Psi^{(n)}  =  \hbox{const}(n)\, t^{-3/2} e^{(n-1/2)(n+3/2)\Omega t} \,k^{n-3/2} \,
 K_0(k/k_{d_n}).
 \label{Psi_n_diff}
\end{equation}

We see that the main effect of the dissipation on all the moments (including the energy spectrum)
is the prefactor changes from $ t^{-1/2} \to  t^{-3/2} $ (but without a change in the exponential growth rate),
further the $K_0(k/k_{d_n})$ form-factor corresponds to a log-correction at $k \ll 1$ and an exponential cut-off at
larger $k$.\footnote{Note that the higher the order of the correlator the earlier the spectral cut-off.}
It is this exponential cut-off that causes the exponential growth of the
mean magnetic energy \cite{kazantsev} and higher $x$-space moments of the magnetic field \cite{chertkov}
to change. Indeed, simply integrating $\Psi^{(n)}$ over $k$ with a cut-off at $k=k_{d_n}$ 
(and ignoring the prefactor and log corrections) one recovers the Kazantsev growth rate of the magnetic energy
in the dissipative regime \cite{kazantsev}. Such an explanation was previously given in \cite{kulsrud}.

\section{Physical interpretation of the scalings}
Let us analyze the physical origins of the scalings obtained in the previous
section because this can give us an indication to whether the same scalings should
be expected for more general strain statistics.

Let us consider the expression (\ref{univ}) for the n-th order correlator and re-write it
in the form $C(n) \cdot exp\big((2n^2+2n) \Omega t) k^{n} \cdot P(k,t)$, where $P(k,t)$
is a universal function. The physical meaning of various terms in this
expression can be easily analyzed. It follows from the Central Limit theorem that 
the large time statistics of the Lyapunov exponents is Gaussian with a dispersion 
$D(t) \sim \sqrt{t}$. It follows from time reversal invariance that the average values of
the Lyapunov exponents are $\bar{\lambda}, 0, -\bar{\lambda}$. Hence, in the large time limit, 
$B(t) \sim exp(\lambda t)$, where $p.d.f. (t) \sim exp(-\frac{(\lambda-\bar\lambda)^2}{2 \Delta t^{-1}})$.
Therefore, the statistics of the magnetic field is log-normal and
$\langle B^{2n} \rangle \sim exp((2\bar{\lambda} n^2+2\Delta n)t)$.
For the statistics chosen in the present paper $\bar{\lambda}
=\Delta=\Omega $. Substituting these values into the last expression for
$\langle B^{2n} \rangle$, we restore the correct
$n$-dependence of the exponential growth rate of $\Psi ^{n}$. Note that terms
of order $n^2$ in the growth rate are due to the Gaussian nature of the fluctuations
of $\lambda$ around $\bar{\lambda}$, while the terms of order $n$ are due to
the fact that $\bar{\lambda} \neq 0$. It is interesting that the magnetic field has
log-normal statistics in both the perfect conductor and the dissipative regimes.
For the coordinate space moments of $B$, persistence of the log-normality was 
emphasized in \cite{schek3}, although the $e^{\lambda n^2 t}$ dependence was 
established earlier in \cite{chertkov}. An equivalent result for the
random matrices can be traced back to \cite{fustenberg} (see also \cite{falkovich}).

The $k$-dependence of the magnetic field correlators is also very important because it
gives us information about the dominant structures in wavenumber space.
Suppose that initially the magnetic turbulence is isotropic and concentrated in a ball
centered at the origin in wavenumber space. For each realization such a ball will
stretch into an ellipsoid with one large, one short and one neutral dimension.
One can visualize this ellipsoid as an elongated flat cactus leaf with thorns aligned to the
direction of the magnetic field. Note that in this picture one component of the magnetic field
(transverse to the cactus leaf) is dominant, this is captured by the fact that the
polarization $W$ introduced in this paper tends to zero at large time.
Another consequence of this picture is that the wavenumber space will be covered by 
the ellipsoids more and more sparsely at large $k$, this implies large intermittent 
fluctuations of the magnetic field Fourier transform. These fluctuations can be quantified 
by the flatness $F$ which was shown in (\ref{Fsoln}) to grow as $ k^{3/2}$, a clear 
indication of the small-scale intermittency. 

\begin{figure}[!Ht]
\begin{center}
\includegraphics[width=.49\linewidth]{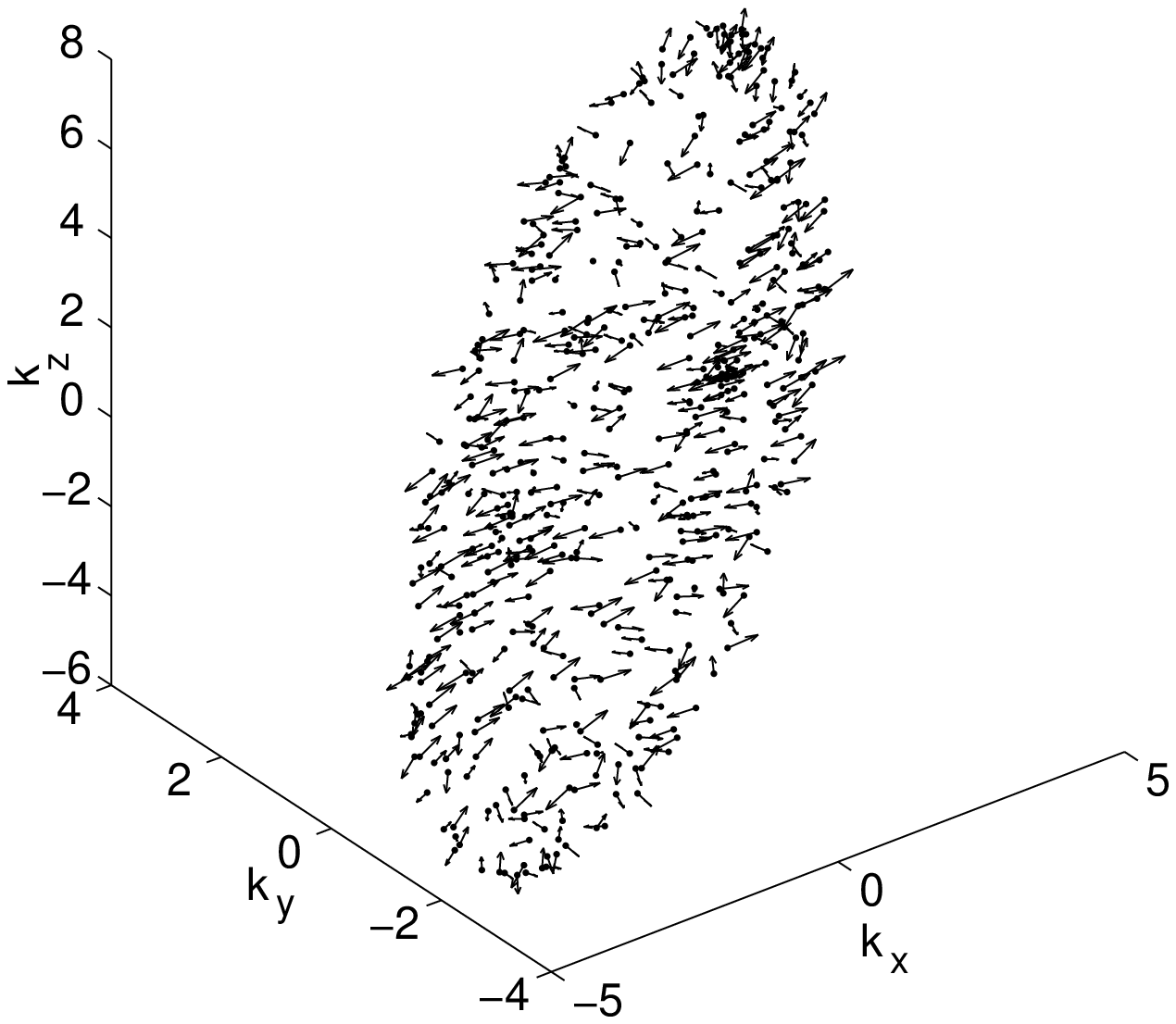}
\includegraphics[width=.49\linewidth]{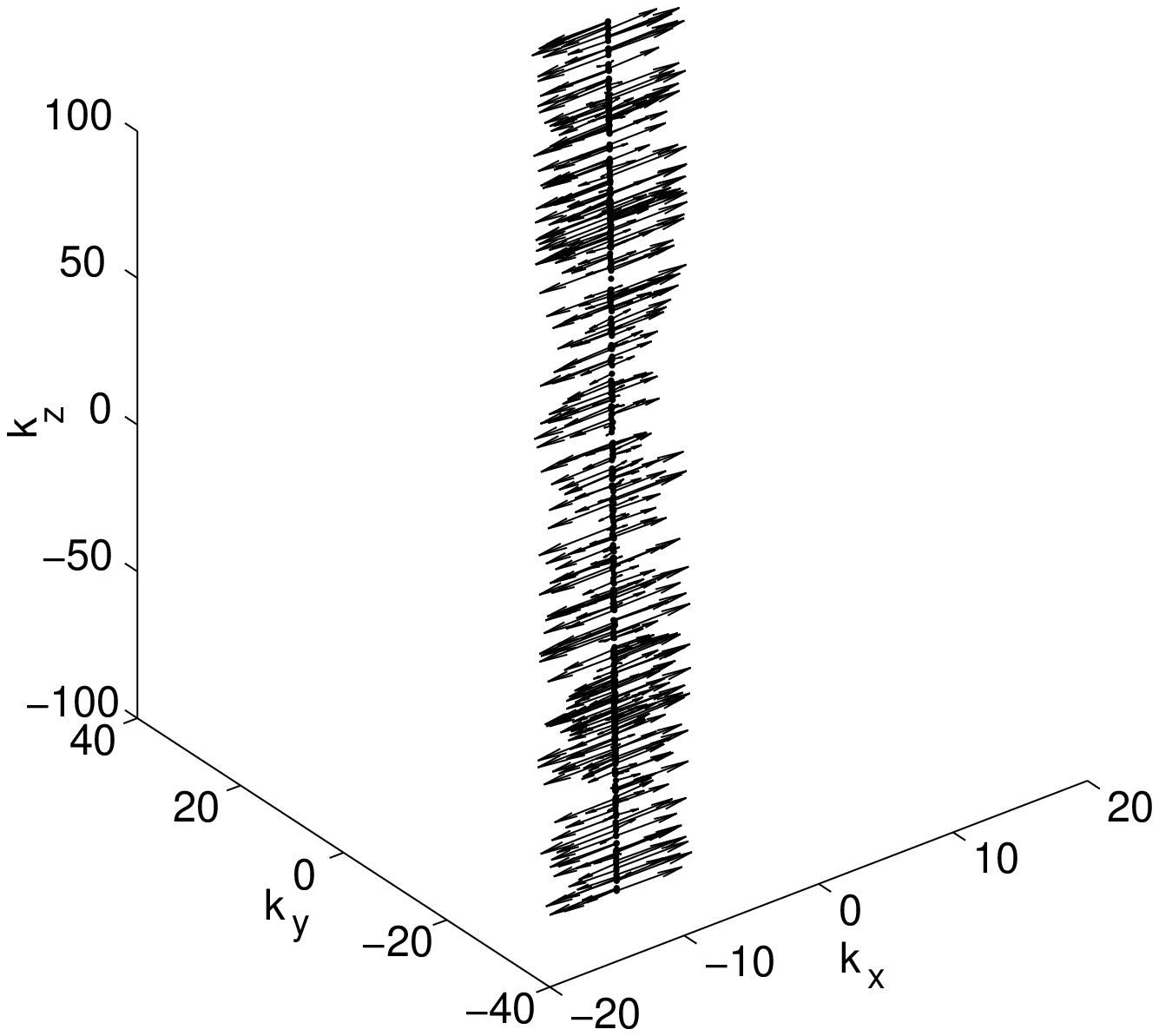}
\caption{The magnetic field of 500 wavepackets 
in a numerical simulation with $\Omega = 0.36$ and 
$\kappa=0$. The figures show the particle's positions in $k$-space
at $t=6$ and their corresponding real magnetic fields. The
left and right figures correspond to different realisations 
of the strain matrix.
}\label{Ellipsoids1}
\end{center}
\end{figure}

\begin{figure}[!Ht]
\begin{center}
\includegraphics[width=.49\linewidth]{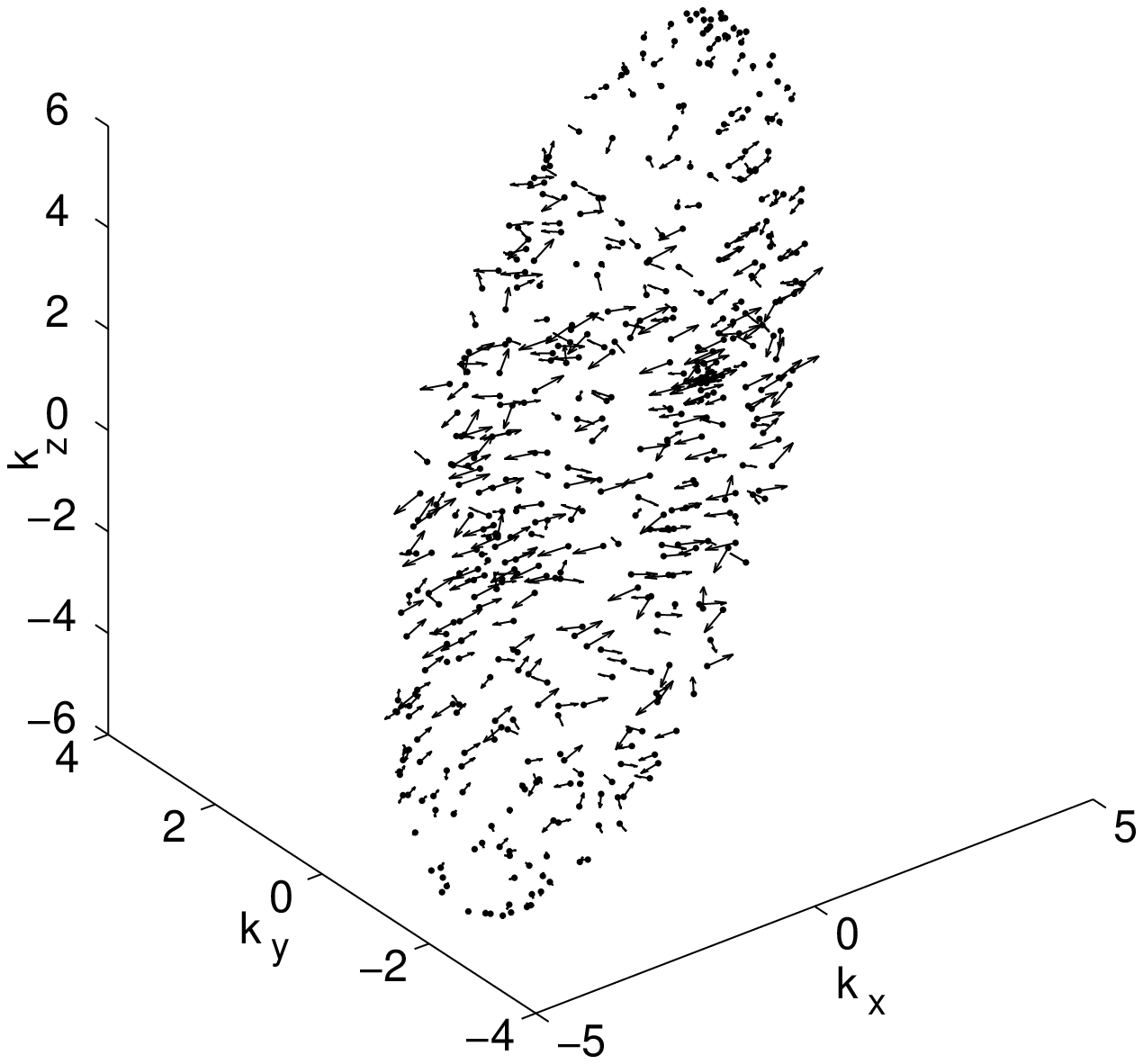}
\includegraphics[width=.49\linewidth]{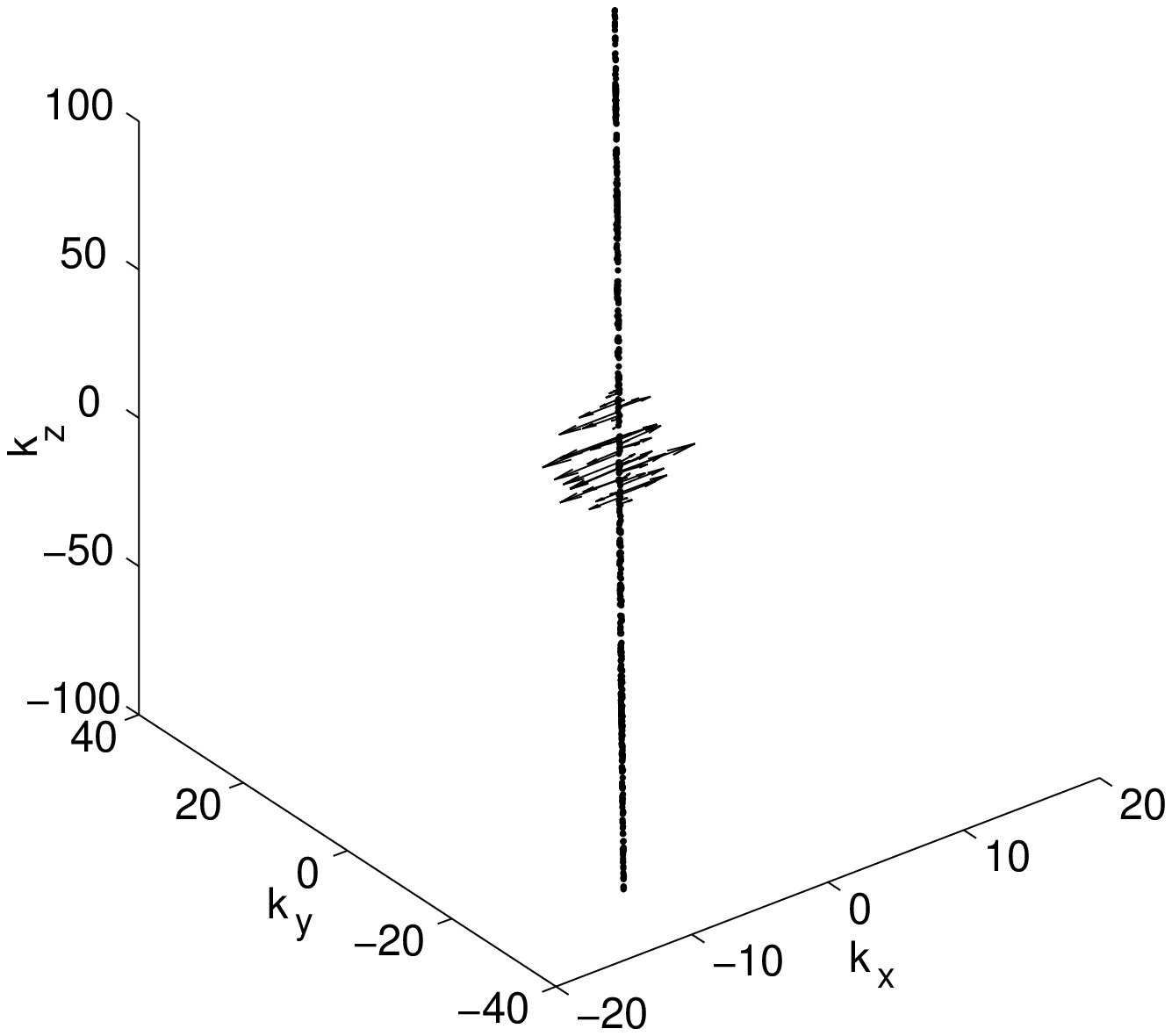}
\caption{The magnetic field of 500 wavepackets 
in a numerical simulation with $\Omega = 0.36$ and 
$\kappa=0.005$. The figures show the particle's positions in $k$-space
at $t=6$ and their corresponding real magnetic fields for 
two different realisations of the strain matrix. The left figure
corresponds to the same strain field as the left-hand graph of 
figure \ref{Ellipsoids1} and similarly for the right-hand graphs.
}\label{EllipDiff}
\end{center}
\end{figure}

To investigate the Kraichnan-Kazantsev model based dynamo problem further, a set of numerical experiments 
have been performed to, among other things, investigate the sensitivity of these 
analytical results to changes in the strain statistics. The details of 
this investigation can be found in \cite{robsPhD,NumPaper}. To help visualise the cactus leaf
structures described above we will briefly include some snapshots of the configuration 
of a wavepacket ensemble in wavenumber space. As above each wavepacket is 
initially randomly distributed on a unit sphere in $k$-space with a randomly orientated 
magnetic field that lies in a plane perpendicular to the wavepacket's wavevector,
tangent to the sphere. Figure \ref{Ellipsoids1} shows the real magnetic fields of a 
set of $500$ wavepackets that have been subjected to two different realisations of 
the strain matrix.\footnote{The imaginary magnetic fields are qualitatively similar 
and have therefore not been included.} In the left-hand figure it is clear the 
magnetic field in this realisation is far from being plane polarised,
the magnetic vectors still being predominantly random in their orientation at this given point in 
time. In contrast, in the right-hand figure, which has also been taken at the same point in time, 
we see the ellipsoid has become very elongated and the magnetic field appears plane polarised.
In contrast, figure \ref{EllipDiff} shows the same two strain realisations but with a finite 
diffusivity. The right-hand figure in particular demonstrates why, in the diffusive regime, 
an ellipsoid will cover $k$-space more sparsely due to the decay of its magnetic field at 
its tips. This is the reason why spectral flatness increases in the diffusive regime.

\section{Conclusion}
In this paper we introduced a description of the small-scale magnetic
turbulence in terms of the one-point correlators of the Fourier amplitudes.
From the classic works of \cite{kazantsev,kraichnan,batchelor} to more recent 
analysis (see for example \cite{schekoch} and references therein), 
the only correlator of this kind that was studied is the energy spectrum.
The higher order correlators have only been considered in coordinate
space \cite{chertkov,falkovich}. In this paper we have considered the Fourier 
correlators of all orders and showed that they contain important additional 
information about the turbulence, which is unavailable from a similar analysis of 
one-point or two-point coordinate space correlators. In particular, the fourth order 
Fourier correlators carry information about the mean polarization of the magnetic 
field modes. We showed that this polarization becomes plane for the Kraichnan-Kazantsev model. 
The $k$-scaling of the higher correlators allows us to determine the structures in Fourier space 
responsible for the intermittency, which for the Kraichnan-Kazantsev dynamo turns out to be 
elongated ellipsoids centered at the origin. The time scaling of the higher correlators allows 
one to conclude that the magnetic field has a log-normal statistics, although the same 
information is contained, and was established before, in analysis of the coordinate space correlators 
\cite{chertkov,falkovich,schek3}.

Finally, we would like to discuss an interesting connection between this work and the recent work 
of Schekochihin {\em et al.} \cite{schek2,schek3}. In particular the connection between the 
statistics of the magnetic field curvature studied by Schekochihin {\em et al.} and the 
magnetic polarization measure introduced in this present paper. 
Schekochihin {\em et al.} found that that the curvature of the magnetic field
decreases, corresponding to folded and strongly stretched structures.
This agrees with our results that the Fourier modes of the magnetic field
tend to a state of plane polarization. However, the polarization
gives more information than the curvature statistics. Indeed, zero curvature
allows any structure that is constant along the magnetic field, in particular
a set of magnetic filaments parallel to each other or a set of layered
sheets, such that the magnetic field is constant on each sheet but
its direction may change arbitrarily when passing from one layer
to another. On the other hand, the wavenumber scalings obtained in this 
paper, (\ref{Psi_n_pc}) and (\ref{Psi_n_diff}), indicate that the magnetic 
field structures are layers in coordinate space and this rules out any 
filamentary structures. Further, our result about the plane polarization inhibit
any ``twists'' of the magnetic field between layers, i.e. the magnetic field
direction stays the same (or reverses) when passing from one layer to another. 
In fact, the presence of one neutral direction in the Lagrangian deformations 
tells us that these layers have a finite width in one direction and thus look 
like ribbons with the magnetic field directed along these ribbons.
 
\appendix
\section*{}

Our aim here is to derive a closed equation for the generating function $Z$
(\ref{zdotFF1}), starting with equation (\ref{zdot}). The last term in this 
equation is the easiest one
\begin{equation}
 - 2 \kappa k^2 \langle (\lambda|{\mathbf B}({\mathbf k})|^2 
 + \alpha {\mathbf B}^2({\mathbf k}) + \beta \overline{{\mathbf B}}^2({\mathbf k})) E
 \rangle = -2 \kappa k^2 {\cal D} Z,
 \label{last}
\end{equation}
where ${\cal D}$ is a differential operator defined in
(\ref{calD}). The correlators containing a factor of $\sigma_{ij}$ can be found
using Gaussian integration by parts. In particular
\begin{eqnarray}
 \langle \sigma_{ij} E \rangle  =  \Omega \langle \frac{\delta E}{ \delta \sigma_{ij} } \rangle =
 \frac{\Omega}{2} 
&&[ \lambda \langle (\Gamma_{m,ij} {\overline B_m} + {\overline \Gamma_{m,ij}} {B_m}) E \rangle \label{g_int} \\
 &&+ 2 \alpha   \langle \Gamma_{m,ij} { B_m}  E \rangle +
 2 \beta  \langle \overline \Gamma_{m,ij} {\overline B_m}  E \rangle ],
 \nonumber
\end{eqnarray}
where we have used the definition (\ref{E}). Here, $\Gamma_{m,ij}$ is a response function
\begin{equation}
 \Gamma_{m,ij} = \frac{\delta B_m}{\delta \sigma_{ij} }.
 \label{gam} 
\end{equation}
Differentiating (\ref{beqn}) with respect to $\sigma_{ij}$  and using the statistical whiteness of the
strain tensor we get
\begin{equation}
 \Gamma_{m,ij} = \left[ k_i \, \partial_j - \frac{\delta_{ij}}{d} (1 + k_l \, \partial_l)\right]  B_m 
 + \delta_{mi} B_j.
 \label{gam1} 
\end{equation}
In what follows we will make use of the isotropy of the turbulence, in particular, expressions of the type
\begin{eqnarray}
  \langle (\overline B_i B_j +  \overline B_j B_i ) E \rangle & =& 
 \frac{2}{d-1}  \langle |B|^2  E \rangle \left(\delta_{ij} - \frac{k_i k_j}{k^2}\right), \\
  \langle B_i  B_j  E \rangle &=&
 \frac{1}{d-1}  \langle B^2  E \rangle \left(\delta_{ij} - \frac{k_i k_j}{k^2}\right), \\
  \langle \overline B_i  \overline B_j  E \rangle &=&
 \frac{1}{d-1}  \langle \overline B^2  E \rangle \left(\delta_{ij} - \frac{k_i k_j}{k^2}\right). 
\end{eqnarray}
Substituting (\ref{gam1}) into (\ref{g_int}) and using the above isotropy relations
we have
\begin{equation}
 \langle \sigma_{ij} E \rangle  =  \frac{\Omega}{2} \left(k_i \partial_j - \frac{\delta_{ij}}{d}
 k_l \partial_l \right) Z - \frac{\Omega}{d-1} \left( \frac{k_i k_j}{k^2} - \frac{\delta_{ij}}{d} \right) {\cal D} Z,
 \label{sije}
\end{equation}
where ${\cal D}$  is a differential operator defined in
(\ref{calD}). This allows us to find the first term on the right hand-side of (\ref{zdot})
\begin{equation}
 k_i \partial_j \langle \sigma_{ij} E \rangle  =  \frac{\Omega}{2} \left[ \frac{d-1}{d} k^2 Z_{kk} +
 \frac{1}{d} \left(-2 {\cal D} + d^2 -1\right) k Z_k -2 {\cal D} Z \right].
 \label{1st}
\end{equation}
Similarly, the other three terms on the right hand-side of (\ref{zdot}) can be
obtained via Gaussian integration by parts, and the use of the
response function (\ref{gam1}) and isotropy condition. After some lengthy but
straightforward algebra one obtains
\begin{eqnarray}
 \lambda \langle \sigma_{ml} ({\overline B_m} B_l + {\overline B_l} B_m) E
 \rangle = \lambda \Omega &&\left[
 \left(d - \frac{2}{d}\right) Z_\lambda + 2 \left(1 - \frac{1}{d}\right) {\cal D} Z_\lambda \label{2nd}\right.\\
 &&\left. + \lambda (Z_{\alpha \beta} - Z_{\lambda \lambda }) -
 \frac{1}{d} k_i \partial_i Z_\lambda \right],
 \nonumber
\end{eqnarray}
and
\begin{eqnarray}
 2 \alpha  \langle \sigma_{ml} { B_m} B_l E \rangle
 =  \alpha \Omega &&\left[
 \left(d - \frac{2}{d}\right) Z_\alpha + 2 \left(1 - \frac{1}{d}\right) {\cal D} Z_\alpha \right. \label{3rd}\\
 &&\left. - 2 \beta Z_{\alpha \beta} + 2 \beta Z_{\lambda \lambda } -
 \frac{1}{d} k_i \partial_i Z_\alpha \right].\nonumber
\end{eqnarray}
The fourth term can be obtained from (\ref{3rd}) via interchanging 
$\alpha $ with $\beta$ and $B$ with $\overline B$
\begin{eqnarray}
 2 \beta  \langle \sigma_{ml} {\overline B_m} \overline B_l E \rangle
 = \beta \Omega &&\left[
 \left(d - \frac{2}{d}\right) Z_\beta + 2 \left(1 - \frac{1}{d}\right) {\cal D} Z_\beta \label{4th} \right.\\ 
 && \left. - 2 \alpha Z_{\alpha \beta} + 2 \alpha Z_{\lambda \lambda } -
 \frac{1}{d} k_i \partial_i Z_\beta \right].
 \nonumber
\end{eqnarray}
Using the expressions (\ref{1st}), (\ref{2nd}), (\ref{3rd}), 
(\ref{4th}) and (\ref{last}), we obtain the final equation
\begin{eqnarray}
 \dot Z &=& \frac{\Omega}{2} \left[ \left(1 - \frac{1}{d}\right) k^2 Z_{kk} + 
 \frac{1}{d} (-4 {\cal D} + d^2 -1) k Z_{k} \right. \label{zdotF1} \\
 && \left. +(2d-6) {\cal D} Z + 4\left(1-\frac{1}{d}\right) {\cal D}^2 Z
 + 2 (\lambda^2 -4 \alpha \beta) (Z_{\alpha \beta}  - Z_{\lambda \lambda})
 \right]
 -2 \kappa k^2 {\cal D} Z.
 \nonumber
\end{eqnarray}



%

\newpage

\end{document}